\PassOptionsToPackage{doi=false,isbn=false,url=false}{bibtex}
\documentclass[letterpaper, 10 pt, conference]{ieeeconf} 

\IEEEoverridecommandlockouts                              

\usepackage{graphicx}
\graphicspath{ {./figures/} }
\usepackage{amsmath}
\usepackage{amssymb}
\usepackage{algorithm}
\usepackage[noend]{algpseudocode}
\usepackage{xcolor}
\definecolor{tumblue}{RGB}{0,101,189}

\usepackage{stfloats}
\usepackage{booktabs}
\usepackage{setspace}
\usepackage{tabularx}
\usepackage{booktabs}
\usepackage[T1]{fontenc}
\usepackage[scaled=0.85]{beramono}
\usepackage{listings}
\usepackage{caption}
\usepackage{cleveref}
\usepackage{url}
\let\labelindent\relax
\usepackage{enumitem}
\usepackage{subfigure}
\usepackage[export]{adjustbox}
\usepackage{multicol}
\usepackage{multirow}
\usepackage{amssymb}
\usepackage{booktabs}
\usepackage{caption}
\usepackage{needspace}

\usepackage{pgfplots}
\usepackage{contour}
\usepackage{comment}
\usepackage{color,soul}
\usepackage{tikz}
\usepackage{algorithm}
\usepackage{algpseudocode}

\newtheorem{defn}{Definition}

\Crefname{lstlisting}{Listing}{Listings}
\lstdefinestyle{sparqlstyle}{
  language=OCL,
  basicstyle=\footnotesize,
  stepnumber=1,
  numbersep=10pt,
  tabsize=2,
  showspaces=false,
  breaklines=true
}

\usepackage[nonumberlist,nopostdot,acronym,toc]{glossaries}

\usetikzlibrary{automata, positioning, arrows.meta,calc,cd}
\usetikzlibrary{patterns}
\usetikzlibrary{datavisualization}
\usetikzlibrary{shapes.arrows}
\usetikzlibrary{patterns.meta}

\tikzdeclarepattern{
  name=hatch,
  parameters={\hatchsize,\hatchangle,\hatchlinewidth},
  bounding box={(-.1pt,-.1pt) and (\hatchsize+.1pt,\hatchsize+.1pt)},
  tile size={(\hatchsize,\hatchsize)},
  tile transformation={rotate=\hatchangle},
  defaults={
    hatch size/.store in=\hatchsize,hatch size=5pt,
    hatch angle/.store in=\hatchangle,hatch angle=0,
    hatch linewidth/.store in=\hatchlinewidth,hatch linewidth=.4pt,
  },
  code={
      \draw[line width=\hatchlinewidth] (0,0) -- (\hatchsize,\hatchsize);
  }
}
\tikzset{
->, 
>=stealth, 
node distance=3cm, 
every state/.style={thick, fill=gray!10}, 
initial text=$ $, 
}
\usepackage{pgfplots}
\pgfplotsset{compat=newest}
\usetikzlibrary{patterns,intersections}
\usepgfplotslibrary{fillbetween}
\usepackage{contour}

\usepackage{xcolor}
\usepackage{marginnote}
\usepackage{comment}

\definecolor{rev1}{RGB}{236,0,140}   
\definecolor{rev2}{RGB}{0,150,0}   

\newenvironment{revisioncolor}[2]{%
    \marginnote{\textcolor{#2}{\textbf{#1}}}[0pt]%
    \color{#2}%
}{%
    \color{black}%
}

\newacronym{cps}{CPS}{Cyber Physical System}
\newacronym{des}{DES}{Discrete Event System}
\newacronym{fsa}{FSA}{Finite State Automata}
\newacronym{ltlf}{LTLf}{Linear Temporal Logic on Finite Traces}
\newacronym{ltl}{LTL}{Linear Temporal Logic}
\newacronym{llm}{LLM}{Large Language Model}
\newacronym{tl}{TL}{Temporal Logic}
\newacronym{nl}{NL}{Natural Language}
\newacronym{mas}{MAS}{Multi-Agent System}
\newacronym{pa}{PA}{Product Agent}
\newacronym{ra}{RA}{Resource Agent}
\newacronym{cca}{CCA}{Central Controller Agent}
\newacronym{dag}{DAG}{Directed Acyclic Graph}
\newacronym{dfs}{DFS}{Depth First Search}
\newacronym{cegis}{CEGIS}{Counterexample Guided Inductive Synthesis}

\usepackage{balance} 
\usepackage{placeins}
\usepackage{float}

\makeatletter
\newcommand\fs@betterruled{%
  \def\@fs@cfont{\bfseries}\let\@fs@capt\floatc@ruled
  \def\@fs@pre{\vspace*{5pt}\hrule height.8pt depth0pt \kern2pt}%
  \def\@fs@post{\kern2pt\hrule\relax}%
  \def\@fs@mid{\kern2pt\hrule\kern2pt}%
  \let\@fs@iftopcapt\iftrue}
\floatstyle{betterruled}
\restylefloat{algorithm}
\makeatother

\author{Jonghan Lim$^{1}$, Mostafa Tavakkoli Anbarani$^{2}$, Rômulo Meira-Góes$^{3}$, and Ilya Kovalenko$^{1,2}$
\thanks{The authors [1] are with the Department of Industrial and Manufacturing Engineering, [2] are with the Department of Mechanical Engineering, and [3] are with the Department of Electrical Engineering, at the Pennsylvania State University, State College, USA
        (e-mail: \{jxl567; mkt5457; romulo; iqk5135\}@psu.edu).
        }%
\thanks{This material is based upon work supported by the National Science Foundation (NSF) under CAREER Award No.~2442362.}
}

\title{\LARGE \bf
Logic-Based Verification of Task Allocation for LLM-Enabled Multi-Agent Manufacturing Systems
}

\begin{document}
\setlength{\textfloatsep}{2pt}

\maketitle
\thispagestyle{empty}
\pagestyle{empty}

\begin{abstract}

Manufacturing industries are facing increasing product variability due to the growing demand for personalized products. Under these conditions, ensuring safety becomes challenging as frequent reconfigurations can lead to unintended hazardous behaviors. Multi-agent control architectures have been proposed to improve flexibility through decentralized decision-making and coordination. However, these architectures are based on predefined task models, which limit their ability to adapt task planning to new product requirements while preserving safety. Recently, large language models have been introduced into manufacturing systems to enhance adaptability, but reliability remains a key challenge. To address this issue, we propose a control architecture that leverages the flexibility of large language models while preserving safety on the manufacturing shop floor. Specifically, the proposed framework verifies large language model-enabled task allocations by using temporal logic and discrete event systems. The effectiveness of the proposed framework is demonstrated through a case study that involves a multi-robot assembly scenario, showing that unsafe tasks can be allocated safely before task execution.

\end{abstract}

\section{Introduction}
\label{section-intro}
Manufacturing systems need to accommodate higher product variety and dynamic operating conditions due to increasing demand for personalized products~\cite{gu2024smart}. These trends increase the need for control strategies that support adaptation of the production system while maintaining reliability~\cite{vogel2015evolution}. To address these challenges, \gls{mas} has been proposed as a manufacturing control paradigm that distributes task execution and resource coordination through interactions among agents representing products and resources~\cite{kovalenko2019model}. \glspl{mas} have been extended by formalizing agent behaviors and interactions using \gls{des} models for structured representation of task execution and resource coordination~\cite{kovalenko2020priced, kovalenko2022cooperative}. These works enable agents to dynamically explore the environment, apply optimization-based planning, and negotiate constraints for adaptive scheduling. These approaches provide structured models for adaptive decision-making, however, they rely on predefined models and decision-making components. They have limited ability to adapt to unforeseen product requirements.

\glspl{llm} have been explored to enhance flexibility in manufacturing systems by enabling adaptive task planning and coordination across heterogeneous resources~\cite{lim2024large, lim2026adaptive}. \glspl{llm} also enable dynamic exploration of resource capabilities in response to unexpected disruptions at runtime~\cite{lim2025exploration}. While these works indicate that \glspl{llm} are effective in improving flexibility beyond predefined scenarios, their outputs are not guaranteed to satisfy safety constraints, such as collision avoidance and precedence relationships. With manufacturing systems increasingly relying on automated software, ensuring correctness is critical since software faults or incorrect state interpretations can lead to hazardous results~\cite{anbarani2024modeling}. These considerations motivate the use of verification in \gls{llm}-enabled \gls{mas} in manufacturing that can improve flexible decision-making while preserving safety.

One approach for verifying automated decision-making in manufacturing is~\gls{tl}. \gls{tl} provides a formal language for specifying temporal constraints on the ordering of events~\cite{pnueli1977temporal}. In prior studies, \gls{tl} has been used to verify the behavior of agents against task specifications. Formalizing \gls{nl} into \gls{tl} enables robot planning and execution under safety constraints \cite{liu2022lang2ltl, xu2025nl2hltl2plan}. However, these approaches focus on planning-level \gls{tl} and do not address verification of generated plans using~\gls{des}, which capture execution-level behaviors. \gls{des} enables explicit representation of events and states, allowing concurrency and resource conflicts to be identified and checked against safety constraints. Other works have explored \gls{llm}-enabled decision-making using \gls{tl} as an external verification layer. For example, \gls{tl}-based planners have been combined with \glspl{llm} to guide task decomposition while providing formal guarantees \cite{wang2023conformal}. Similarly, constraint-based frameworks enforce safety rules to mitigate errors from language-based reasoning \cite{yang2024plug}. 
While these studies demonstrate that \gls{tl} can constrain \gls{llm}-driven decision-making, they do not address verification in manufacturing involving interacting resources, parts, and constraints. In these environments, safety is determined by system-level execution under shared-resource, synchronization, and ordering requirements. Even when individual tasks are feasible for a single agent, the combined plan of multi-agent settings can be unsafe due to resource conflicts or newly introduced inter-agent dependencies. This motivates the use of \gls{des}-based verification.

In this study, we present a control architecture that enables the verification of \gls{llm}-generated task plans in~\glspl{mas}. The proposed framework is inspired by \gls{cegis} \cite{solar2008program}, where candidate solutions are iteratively checked and revised based on formally detected violations. 
In manufacturing, new product requirements and the violations they introduce cannot always be addressed using predefined models. In addition, a detected violation does not uniquely determine how the plan should be revised. Therefore, we use the verification result as structured feedback to guide flexible replanning.
While \gls{cegis} focuses on bounded synthesis, we utilize~\glspl{llm} to translate new product requirements into multi-agent task plans by composing resource capabilities through reasoning over requirement-capability relationships. \gls{des} provides execution-level details for these plans, enabling them to be formally evaluated. This way, logic-based verification ensures correctness without redesigning the models. The main contributions of this work are: (1) a control architecture that supports verification of \gls{llm}-generated task plans, (2) a verification process that evaluates task plans with~\gls{tl} represented as automata, and (3) a feedback mechanism that utilizes verification results to refine task plans. The rest of this paper is organized as follows. Section~\ref{section-preliminaries} introduces the preliminaries on~\gls{tl} and~\gls{des}. Section~\ref{section-overview} describes the formulation of the architecture for verifying \gls{llm}-generated task plans. Section~\ref{section-verification} presents the \gls{tl}-based verification process and its automata. Section~\ref{section-case study} illustrates the approach through a case study. Finally, the conclusion is given in Section~\ref{section-conclusion}.


\section{Preliminaries}
\label{section-preliminaries}
\subsection{Finite State Automata}
\label{subsec:fsa}

\gls{fsa} is a modeling framework in~\gls{des} and used to express execution logic and process sequencing~\cite{lafortune:2021}. Manufacturing system behavior is represented using~\gls{fsa} to support verification of temporal safety constraints. A finite-state automaton is defined as $\mathcal{A} = (X, E, Tr, x_0, X_m)$, where $X$ is a finite set of states, $E$ is a set of events, $Tr : X \times E \rightarrow X$ is a transition function, $x_0 \in X$ is the initial state, and $X_m \subseteq X$ is the set of marked states corresponding to desired behaviors~\cite{lafortune:2021}.

\subsection{Linear Temporal Logic on Finite Traces}

A major part of the safe execution of plans is characterized by satisfaction of temporal constraints. A task plan is considered safe depending on the ordering and occurrence of tasks. Even when a task plan is logically correct, incorrect ordering or conflicting use of shared resources can propagate fault sequences. For this reason, temporal constraints must be enforced. In the proposed approach, \gls{ltlf}~\cite{de2013linear} is used to specify constraints. Corresponding formulas are used to evaluate over finite execution traces, composed of discrete events. To formally specify temporal constraints, we define atomic propositions that characterize observable events.

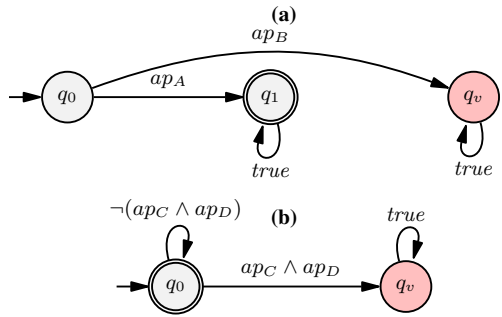
\begin{figure}[t]
\centering
\scalebox{0.9}{
\begin{tikzpicture}[
  on grid,
  >={Stealth[inset=0pt,length=9pt,angle'=28,round]},
  state/.style={circle, draw, minimum size=0.75cm, thick, fill=gray!10},
  every edge/.style={draw, thick, font=\small},
  every node/.style={font=\small}
]

\node at (3.2,1.2) {\textbf{(a)}};

\node[state, initial] (q0a) at (0,0) {$q_0$};
\node[state, accepting, double] (q1a) at (3.0,0) {$q_1$};
\node[state, fill=red!25] (qda) at (6.0,0) {$q_v$};

\draw (q0a) edge[above] node {$ap_A$} (q1a);
\draw (q0a) edge[bend left=20, above] node {$ap_B$} (qda);

\draw (q1a) edge[loop below] node {$\mathit{true}$} (q1a);
\draw (qda) edge[loop below] node {$\mathit{true}$} (qda);

\node at (3.2,-1.8) {\textbf{(b)}};

\node[state, initial, accepting, double] (q0b) at (1.6,-2.8) {$q_0$};
\node[state, fill=red!25] (qdb) at (5.0,-2.8) {$q_v$};

\draw (q0b) edge[loop above] node {$\neg(ap_C \land ap_D)$} (q0b);
\draw (q0b) edge[above] node {$ap_C \land ap_D$} (qdb);

\draw (qdb) edge[loop above] node {$\mathit{true}$} (qdb);

\end{tikzpicture}
}
\caption{Finite-state automata encoding \gls{ltlf} safety constraints:
(a) ordering and (b) mutual exclusion.}

\label{fig:preliminary-example}
\end{figure}

\begin{figure*}[t]
\smallskip
\smallskip
    \centering
    \captionsetup{belowskip=-15pt}
    \includegraphics[width=.98\textwidth]{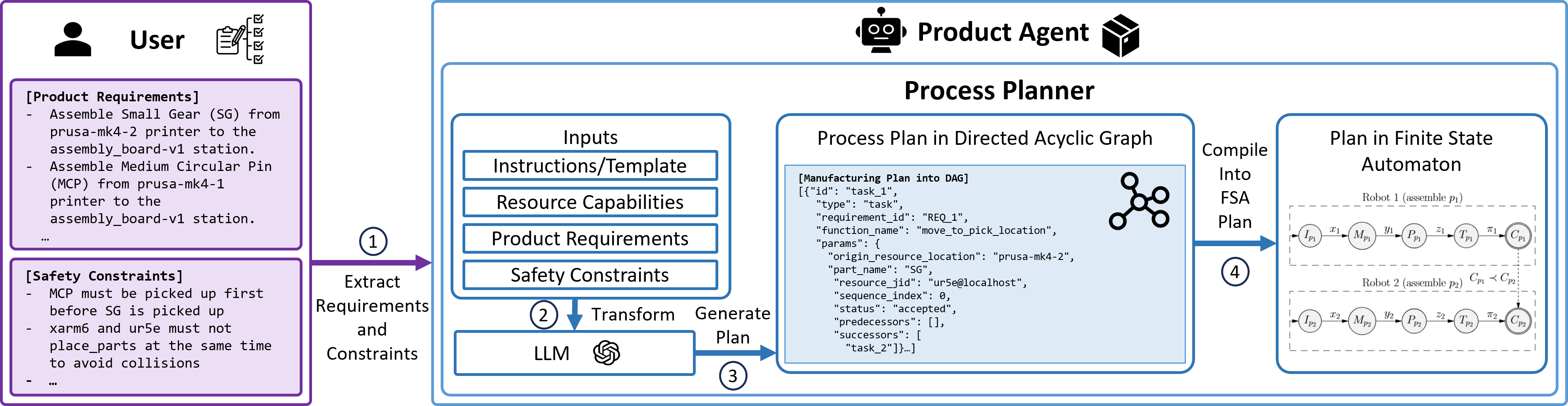}
    \caption{Process planning workflow. \gls{nl} product requirements are transformed into structured requirements using an \gls{llm}, generating a task plan in DAG, and compiling this DAG into a \gls{fsa} plan.}
    \label{fig:planning-workflow}
\end{figure*}

\begin{defn}[Atomic Propositions]
Let $AP$ be a finite set of atomic propositions and $ap \in AP$ be an atomic proposition. In this work, each atomic proposition $ap$ represents the observation of a task execution event and is instantiated using the structured form:
\vspace{-5pt}
\[
\texttt{ap/process/product/resource/event/context}.
\]
\end{defn}

The structured representation follows product, process, resource modeling conventions in manufacturing ontologies~\cite{lemaignan2006mason}. This representation enables uniform specification of task execution events. For example, an $ap$ for an assembly pick operation that has occurred can be expressed
\texttt{ap/assembly/part\_A/robotarm1/pick\_part/printerX}.\\ The \texttt{event} field represents the type of action being observed (e.g., \texttt{pick\_part}, \texttt{place\_part}), while the \texttt{context} field captures relevant conditions, such as location. Using these $ap$ representations, temporal constraints can be specified as \gls{ltlf} formulas. 

\begin{defn}[\gls{ltlf} Formula]
An \gls{ltlf} formula $\varphi$ is constructed from atomic propositions in $AP$ using boolean and temporal operators according to the standard \gls{ltlf} syntax:
\[
\varphi := \top \mid ap \mid \neg \varphi \mid \varphi_1 \land \varphi_2 \mid X \varphi \mid \varphi_1 \, U \, \varphi_2,
\]
where $ap \in AP$. The symbol $\top$ is the boolean constant \emph{true}, $\neg$ and $\land$ is boolean negation and conjunction, respectively. $X$ and $U$ are the \emph{next} and \emph{until} temporal operators.
The boolean operators $\lor$ and $\rightarrow$, as well as the temporal operators \emph{eventually} ($F$) and \emph{globally} ($G$), are also used. They express disjunction, implication, eventual occurrence, and conditions that must hold over the execution trace, respectively.
\end{defn}

Each safety constraint is specified as an~\gls{ltlf} and evaluated over finite execution traces. For example, an ordering constraint can be expressed as $\varphi=\neg ap_B\,U\,ap_A$, which specifies a precedence relationship between task execution events. Another example is a mutual exclusion, $\varphi = G\,\neg(ap_C \land ap_D)$, which enforces that two observable events do not occur simultaneously. These constraints can be translated into equivalent \gls{fsa} for trace-based validation, as shown in Fig.~\ref{fig:preliminary-example}.

\begin{defn}[Safety Automaton]
Each temporal safety constraint, which is specified as an \gls{ltlf} in \textit{Definition 2}, is translated into a safety automaton $\mathcal{A}_s = (Q_s, \Sigma_s, \delta_s, q_{0,s}, Q_{v,s})$, 
where $Q_s$ is a finite set of safety states, $\Sigma_s$ is the set of $AP$ in the safety specification, $\delta_s : Q_s \times 2^{\Sigma_s} \rightarrow Q_s$ is the transition function, $q_{0,s} \in Q_s$ is the initial state, and $Q_{v,s} \subseteq Q_s$ is the set of violation states. A safety constraint is violated when a state in $Q_{v,s}$ is reached.
\end{defn}

This definition formalizes the translation of \gls{ltlf}-based safety constraints into automata for verification. By representing each constraint as a safety automaton with violation states, safety verification can be performed, as described in Section~\ref{section-verification}.

\section{Agent and Problem Formulation}
\label{section-overview}
\subsection{Multi-Agent Architecture Overview}

Various \gls{mas} architectures introduce a \gls{pa} and \gls{ra} structure. \glspl{pa} manage processing sequences for physical parts, while \glspl{ra} execute tasks on resources based on coordination with \glspl{pa}~\cite{kovalenko2019model}. In this section, we detail the roles of \gls{pa} and \gls{ra} and introduce the formal representations required for verifying task planning and allocation.

\subsubsection{Resource Agents}

Each \gls{ra} is modeled as a \gls{fsa} capturing the executable behavior of the resource~\cite{lim2025exploration}. The \gls{ra} behavior is modeled as $\mathcal{A}_{r} = (X_r, E_r, Tr_r, x_{0,r}, X_{m,r})$, where \(X_r\) is a set of physical states, \(E_r\) is a set of task execution events that corresponds to a physical actions (e.g., the start or completion of tasks), and \(Tr_r\) maps states and events to resulting states. \(x_{0,r}\) is the initial state, and \(X_{m,r}\) corresponds to task completion for a part. 

\subsubsection{Product Agents}
\label{subsubsec:product-agents}

A \gls{pa} is responsible for realizing the production requirements of a physical product~\cite{kovalenko2019model}. In this work, \gls{llm}-enabled \glspl{pa} interpret \gls{nl} product requirements and available resource capabilities to generate a task plan represented as a \gls{dag}. This representation captures task dependencies and concurrency and supports compilation into an automaton for discrete-event modeling and verification, as illustrated in Fig.~\ref{fig:planning-workflow}. The automaton is formally defined as $\mathcal{A}_{\text{plan}} = (X_{\text{plan}}, E_{\text{plan}}, Tr_{\text{plan}}, x_0, X_m)$, which models the joint behavior of all \glspl{ra} required to satisfy the product requirements. Let $\mathcal{J}$ be the index set of \glspl{ra}. The global state space is defined as $X_{\text{plan}} \subseteq \prod_{j \in \mathcal{J}} X_j$, where $X_j$ denotes the set of physical states of \gls{ra} $j$. Each global state $x \in X_{\text{plan}}$ represents the joint physical state of all resources. Let $G = (T, D)$ be the task plan \gls{dag}, where
$T = \{\tau_1, \dots, \tau_N\}$ is the set of subtasks to satisfy a product requirement, and $D$ is the precedence relation between tasks. From this task set, the event set is defined as
$
E_{\text{plan}} =
\{\,
\tau_i^{\text{start}},\;
\tau_i^{\text{done}}
\mid
\tau_i \in T
\,\},
$
where $\tau_i^{\text{start}}$ and $\tau_i^{\text{done}}$ represent the start and completion of task $\tau_i$, respectively. 
The transition $Tr_{\text{plan}} \subseteq X_{\text{plan}} \times E_{\text{plan}} \times X_{\text{plan}}$ is derived from \gls{dag}. A transition labeled by
$\tau_i^{\text{start}}$ occurs once all predecessor tasks of $\tau_i$ have completed. The initial state represents no tasks executed, and marked states correspond to successful completion of the global task plan.

\subsection{Problem Formulation}

We consider an \gls{llm}-enabled multi-agent manufacturing system that supports adaptive task planning and allocation. Product and safety requirements are provided in \gls{nl} to guide the task planning process. However, due to known limitations of \gls{llm}, the \gls{pa} may misinterpret instructions during task planning that involve concurrency of resources~\cite{lim2024large}. As a result, the synthesized plan may still violate safety constraints, such as incorrect ordering of tasks or unsafe simultaneous actions. The problem addressed in this work is as follows. When new product requirements are introduced, the challenge is to generate task plans without relying on predefined task models (e.g, fixed task sequences) while satisfying all safety constraints on the shop floor.



\subsubsection{Assumptions}

To address this problem, the proposed framework operates under the following assumptions.

\begin{list}{}{\setlength{\leftmargin}{0.7cm}
              \setlength{\labelwidth}{1.0cm}
              \setlength{\itemindent}{-0.0cm}}

\item[A.1] The synthesized formal safety specifications accurately capture the intended safety constraints expressed in \gls{nl} and are validated by a domain expert.

\item[A.2] Task execution events relevant to safety constraints are observable and can be mapped to atomic propositions.

\item[A.3] Product requirements expressed in \gls{nl} are achievable with the given manufacturing capabilities in the system.

\end{list}


These assumptions are introduced to focus on the safety verification problem. A.1 allows the framework to focus on verifying formal constraints rather than semantic correctness. A.2 ensures that safety-relevant events can be observed within the framework. A.3 excludes infeasible planning cases that do not reflect safety-related failures. These assumptions allow the study to focus on verifying the safety properties of generated task plans.

\section{Control Architecture for Verification}
\label{section-verification}
In this section, we introduce the \gls{cca}. Following the control architecture proposed by \cite{kovalenko2022toward}, the \gls{cca} operates with a global view of the \gls{mas} to preserve safety guarantees while supporting adaptive task allocation.
While distributed runtime verification~\cite{francalanza2018runtime} can represent new global constraints, they require those constraints to be translated into local conditions for the affected agents. Since synthesized plans can introduce new dependencies between agents, the \gls{cca} was proposed to verify the joint plan directly.

\subsection{Safety Specification}
\label{subsec:specification}

\begin{figure}[t]
\smallskip
\smallskip
    \centering
    \captionsetup{belowskip=0pt}
    \includegraphics[width=.54\textwidth]{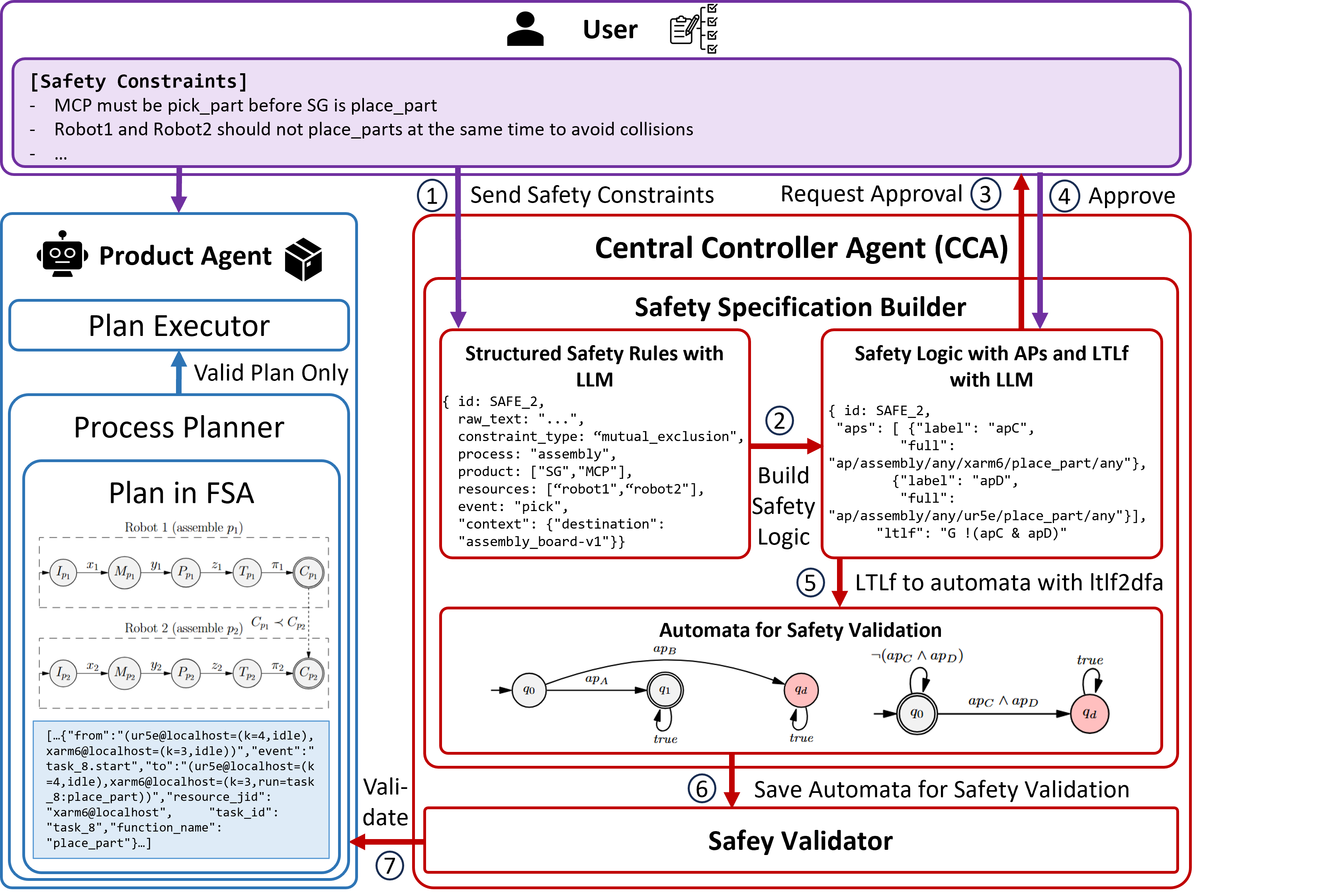}
    \caption{Safety-aware task planning workflow. The \gls{cca} translates \gls{nl} safety constraints into \gls{ltlf} specifications, which are then compiled into \gls{fsa}. These automata are used to validate the \gls{llm}-generated task plan before execution.}
    \label{fig:safety-logic}
\end{figure}

\begin{figure*}[t]
\smallskip
\smallskip
    \centering
    \captionsetup{belowskip=-18pt}
    \includegraphics[width=.95\textwidth]{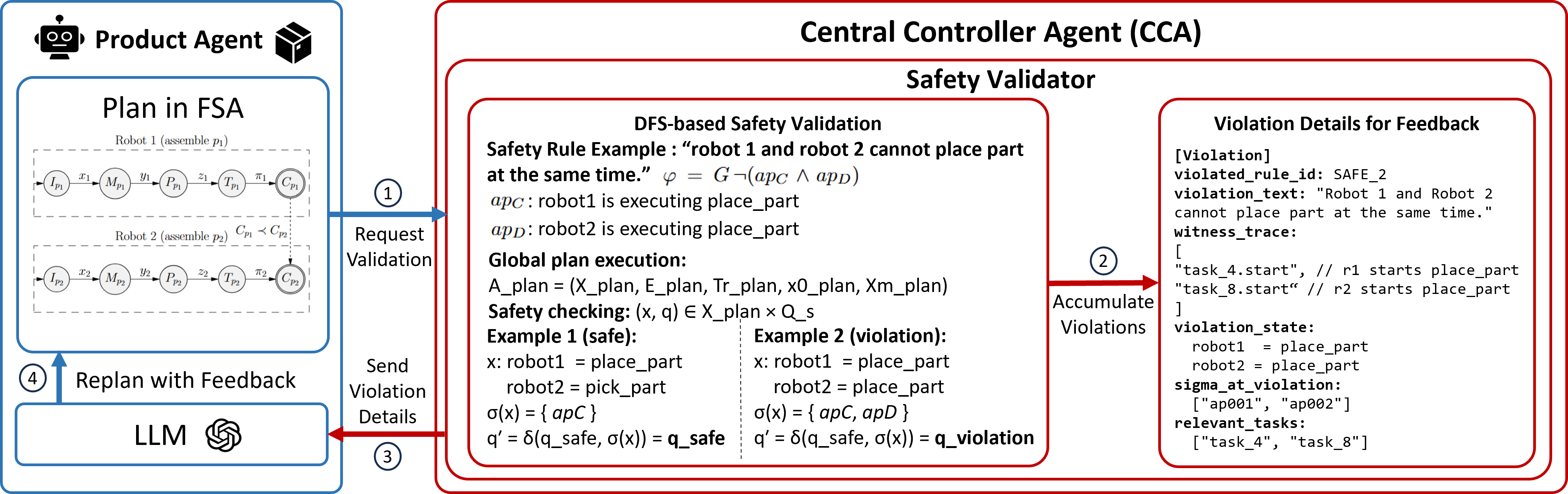}
    \caption{\gls{cca} safety validation architecture. The task plan is traversed with the safety automata using DFS-based validation to detect violations. Safety violations are accumulated and returned as structured feedback for replanning.}
    \label{fig:safety-validator}
\end{figure*}


To enforce safety while accommodating evolving product requirements, we propose a control architecture that formalizes \gls{nl} safety requirements into automata for \gls{des}-based verification, as illustrated in Fig.~\ref{fig:safety-logic}. Safety requirements are specified by a human operator with knowledge of manufacturing safety rules and operational constraints. Let $\mathcal{S} = \{ s_1, \ldots, s_M \}$ be the set of \gls{nl} safety requirements. Each requirement $s_i \in \mathcal{S}$ is transformed into a structured safety representation that includes constraint type, associated process, product, resource, event, and contexts using \gls{llm} to map \gls{nl} safety requirements into structured elements. This representation allows \gls{cca} to convert it into a corresponding set of atomic propositions $AP_i = \{ ap_{i,1}, \ldots, ap_{i,k_i} \} \subseteq AP$, where each \( ap_{i,j} \) represents the observation of a task execution event, as formalized in Definition~1. This set of atomic propositions \( AP_i \) is then used to construct an \gls{ltlf} formula $
\varphi_i \in \mathrm{LTL}_f(AP_i)$, which encodes the safety constraint \( s_i \), as specified in Definition~2. Each formula $\varphi_i$ is translated into a safety automaton $\mathcal{A}_s$ according to Definition~3. 
Since the translation of \gls{nl} safety requirements into formal specifications is not guaranteed, new safety constraints require expert validation~\cite{yang2024plug}. Therefore, based on assumption~A.1, the synthesized \gls{ltlf} and corresponding $\mathcal{A}_s$ are validated by a domain expert to correctly capture the intended safety constraints. As this manual validation limits scalability in highly dynamic environments, future work will explore methods such as fine-tuning to reduce reliance on expert intervention.

\begin{algorithm}
\caption{DFS-Based Safety Validation}
\label{alg:dfs_safety}
\begin{algorithmic}[1]
\footnotesize
\Procedure{ValidateSafety}{$\mathcal{A}_{plan},\,\mathcal{A}_{safety}$}
    \State $(X_{plan}, E, Tr, x_0, X_m) \gets \mathcal{A}_{plan}$
    \For{\textbf{each} safety automaton $\mathcal{A}_{s} \in \mathcal{A}_{safety}$}
        \State $\mathcal{A}_{s} = (Q_s, \Sigma_s, \delta_s, q_{0,s}, Q_{v,s})$
        \State $R \gets \{(x_0, q_{0,s})\}$ \Comment{reachable product states}
        \State $V \gets \emptyset$ \Comment{visited product states}
        \While{$R \neq \emptyset$}
            \State $(x, q) \gets \textbf{Pop}(R)$
            \If{$(x, q) \in V$} \textbf{continue} \EndIf
            \State $V \gets V \cup \{(x, q)\}$
            \If{$x \in X_m$}
                \State $q_{\text{end}} \gets \delta_s(q, \emptyset)$
                \If{$q_{\text{end}} \in Q_{v,s}$}
                    \State \Return \textbf{Violation Found}
                \EndIf
            \EndIf

            \For{\textbf{each} transition $(x, e, x') \in Tr$ \textbf{from} $x$}
                \State $\sigma \gets \textsc{ExtractAPs}(x')$
                \State $q' \gets \delta_s(q, \sigma)$
                \If{$q' \in Q_{v,s}$}
                    \State \Return \textbf{Violation Found}
                \EndIf
                \State $R \gets R \cup \{(x', q')\}$
            \EndFor
        \EndWhile
    \EndFor
    \State \Return \textbf{Safe}
\EndProcedure
\end{algorithmic}
\end{algorithm}

\subsection{Safety Validation}

After the \gls{cca} constructs safety logic from user-defined constraints, it performs safety validation using a reachability analysis. This analysis is conducted on the product of the plan execution automaton $\mathcal{A}_{plan}$ and a set of safety automata $\mathcal{A}_{safety}$ converted from \gls{ltlf}-based temporal safety specifications.
The goal is to determine whether any task plan in an automata violates a safety constraint before execution. The safety validation workflow, performed by \gls{cca}, is illustrated in Fig.~\ref{fig:safety-validator}.  

Safety validation begins when a \gls{pa} submits a compiled plan execution automaton $\mathcal{A}_{plan}$ to the \gls{cca}. The automaton captures the joint execution states of all resources in processing a product. Given $\mathcal{A}_{plan}$ and a set of safety automata $\mathcal{A}_{safety}$, Algorithm~\ref{alg:dfs_safety} details the reachability analysis performed over product states $(x, q)$. 


Starting from the initial product state $(x_0, q_{0,s})$, the \gls{cca} performs a \gls{dfs}-based reachability analysis over product states $(x, q)$. For each plan transition $(x, e, x') \in Tr$, safety atomic propositions are evaluated at the resulting state $x'$. These propositions are used as input to the safety automaton, which transitions to $q' = \delta_s(q, \sigma(x'))$. If $q' \in Q_{v,s}$, a safety violation is reported. When the reachability analysis ends without reaching any violation state $q_{\text{violation}}$, including marked plan states $x \in X_m$, the plan is considered safe. All detected safety violations are accumulated by the \gls{cca} and used as feedback for replanning via \gls{pa}, as described in Section \ref{subsec:replanning}.

\subsection{Replanning with Feedback}
\label{subsec:replanning}

In the proposed framework, replanning with feedback is used to revise an unsafe task plan into a safety-compliant one when the initial plan violates the specified safety constraints. Deterministic repair from a violation trace is effective in common cases, but it remains limited to predefined models. In manufacturing settings, however, a valid revision may depend on the execution context, including resource availability and task progress. Therefore, we use the \gls{llm} to interpret this context and propose revised task plans. For replanning, the \gls{pa} receives a structured feedback prompt, as shown in~\Cref{tab:replanning_prompt}. It includes the unsafe task plan, the violated safety condition, and a witness trace showing how the violation occurs. It also provides the constraints needed to produce a valid repair, including repair instructions, execution context, and an output schema. This structure supports the model to return a feasible plan that can be applied to the system. The replanning loop is bounded by a fixed number of attempts. A formal analysis of convergence and methods to improve convergence reliability is left for future work.

\begin{table}[t]
\centering
\footnotesize
\caption{Representative examples of the feedback prompt.}
\label{tab:replanning_prompt}
\begin{tabular}{p{0.23\linewidth} p{0.66\linewidth}}
\toprule
\textbf{Prompt section} & \textbf{Representative examples} \\
\midrule

System instructions &
You are a plan expert in pre-execution mode. Fix the unsafe plan before execution begins. ... \\

Unsafe plan &
\{id: SG\_MOVE, funcion: move\_loaded, part: SG, resource\_jid: xarm6, predecessors: [SG\_PICK], ...\} \\

Safety violation &
MCP\_MOVE must occur before SG\_MOVE. ... \\

Violation trace &
witness\_events = [SG\_MOVE.start, ...]; the trace shows SG\_MOVE begins before the required MCP\_MOVE. \\

Execution context &
functions = [pick\_part, move\_loaded, ...]; agents = [xarm6, ur5e]; locations = [prusa-mk3, assembly\_board, ...] \\

Output schema &
\{tasks: [\{id: TASK\_ID, predecessors: [TASK\_ID], change\_reason: explanation of the repair\}]\} \\

\bottomrule
\end{tabular}
\end{table}

\section{Case Study}
\label{section-case study}
\subsection{Case Study Setup}

To evaluate the proposed framework, we consider three assembly scenarios with increasing numbers of robots, parts, and rules, where each rule represents a safety constraint. Figure~\ref{fig:case-study} shows Scenario S1 as the representative setup, where robots transport printed parts from source locations to a shared assembly station. In each scenario, the product requirements and safety rules are provided in~\gls{nl}, and the \gls{pa} generates both the task plan and allocation from these inputs. The pure \gls{llm} baseline uses only the~\gls{nl} safety constraints during plan generation. The proposed framework uses the same inputs together with formal verification. The safety constraints are translated into \gls{ltlf}, converted to safety automata, and checked against the generated plan. Scenarios S2 and S3 increase coordination complexity and the joint state space. Across S1, S2, and S3, the scenario scale increases from 2/3/2 to 3/4/3 and 4/6/4 in robots, parts, and safety rules, respectively. The safety rules include both ordering and mutual-exclusion constraints: S1 includes one of each, S2 includes two ordering and one mutual-exclusion rule, and S3 includes two ordering and two mutual-exclusion rules. Each scenario was evaluated over twenty trials, with the replanning loop bounded to a maximum of five repair attempts per trial. GPT-5 by OpenAI was used in this case study~\cite{singh2025openai}.



\subsection{Case Study Results and Analysis}

\begin{table}[t]
\centering
\caption{Evaluation results across three scenarios.}
\label{tab:results}
\setlength{\tabcolsep}{3pt}
\small
\begin{tabular}{lccccc}
\toprule
\multirow{2}{*}{Scen.} & \multirow{2}{*}{\shortstack{Robots/\\Parts/\\Rules}} & \multirow{2}{*}{\shortstack{Pure LLM\\Rule Sat. (\%)}} 
& \multicolumn{3}{c}{Proposed Framework} \\
\cmidrule(lr){4-6}
 &  &  & \shortstack{Rule Sat.\\(\%)} & \shortstack{Repair\\Attempts} & \shortstack{Verif. Time\\(s)} \\
\midrule
S1 & 2/3/2 & 50.00 & 92.50 & 1.80  & 0.10 \\
S2 & 3/4/3 & 75.93 & 91.67  & 2.20 & 1.53 \\
S3 & 4/6/4 & 50.00 & 86.25  & 3.90 & 25.46 \\
\bottomrule
\end{tabular}
\end{table}

The proposed framework improved safety-rule satisfaction over the pure \gls{llm} approach, as shown in~\Cref{tab:results}. We measured rule satisfaction as the percentage of safety constraints defined in the setup that were satisfied by the generated task plan. Rule satisfaction increased from 50\% to 92.5\% in S1, from 75.93\% to 91.67\% in S2, and from 50\% to 86.25\% in S3. The average number of repair attempts increased from 1.8 in S1 to 2.2 in S2 and 3.9 in S3, while the average verification time increased from 0.1s to 1.53s and 25.46s.

These results show that using only \gls{nl} safety constraints did not consistently generate safety-compliant plans. The proposed framework improved safety compliance in all three scenarios. However, the rule satisfaction still remained below 100\% because the \gls{llm} had to interpret violation details and replanning feedback to produce a corrected plan. The results also show that repair and verification cost increase with problem scale. The number of product states explored during verification increased from 170 in S1 to 1,378 in S2 and 14,618 in S3. Larger scenarios introduce more robots and possible parallel task executions, which makes the verification model much larger. Because the safety validator repeats this search for each safety rule, the number of explored states increases rapidly, leading to the large verification time in S3. To improve scalability, the framework will be extended with symbolic and compositional verification methods for larger systems~\cite{malik2023survey}.

\begin{figure*}[t]
\smallskip
\smallskip
    \centering
    \captionsetup{belowskip=-18pt}
    \includegraphics[width=.98\textwidth]{figures/figure05-casestudy.png}

    \caption{Case study for Scenario S1 and its task-planning results. S2 and S3 extend this setup with more robots, parts, and rules.
    (a) Setup for S1, where two robots assemble a small gear (SG) and a medium circular pin (MCP). 
    (b) Task-planning automata that shows safety violations.
    (c) Repaired automata, where the safety constraints are enforced.
    $I$, $M$, $P$, $T$, and $C$ denote idle, move-to-pick, pick, transport, and place states, respectively. $x$, $y$, $z$, and $\pi$ denote the corresponding events.}
 
    \label{fig:case-study}
\end{figure*}

\section{Conclusion}
\label{section-conclusion}
In this work, we present a control architecture for verifying task allocation in \gls{llm}-enabled multi-agent manufacturing systems. Specifically, the proposed framework utilizes an \gls{llm}-enabled \gls{pa} to dynamically generate a task plan with a \gls{cca} that enforces safety constraints through automata-based verification. By structuring \gls{nl} safety constraints into \gls{ltlf} specifications and corresponding \gls{fsa}, the architecture validates the task plans before task execution. An assembly case study scenario demonstrates that unsafe task allocations can be formally detected and repaired, without limiting planning flexibility. Future work will focus on improving the robustness and generalizability of the proposed verification framework across diverse manufacturing scenarios. 
As the framework scales to larger manufacturing plants, a single \gls{cca} can become impractical, motivating the use of subsystem-level or hierarchical verification architectures to reduce communication and computation bottlenecks.
The framework will also be extended to enable online safety verification to support runtime adaptation when unexpected events occur while preserving safety constraints.

\balance


\end{document}